\documentclass[aps,preprint,preprintnumber,nofootinbib,amsmath,amssymb,ascmac,bm,12pt]{revtex4}
\usepackage{bm}
\usepackage{graphicx}
\usepackage{dcolumn}
\usepackage{bm}

\usepackage{color}

\newcommand{\epzero}{\varepsilon_0}
\newcommand{\kB}{k_{\rm B}}

\newcommand{\mpr}{m_{\rm p}}
\newcommand{\mel}{m_{\rm e}}

\newcommand{\Pp}{P_{\rm p}}
\newcommand{\Pe}{P_{\rm e}}

\newcommand{\rg}{r_{\rm g}}

\newcommand{\vomg}{v_\Omega}

\newcommand{\qE}{q_{\rm BH}}
\newcommand{\hqE}{\hat{q}_{\rm BH}}
\newcommand{\qmin}{q_{\rm min}}
\newcommand{\qmax}{q_{\rm max}}

\newcommand{\calT}{{\cal T}}
\newcommand{\calE}{{\cal E}}
\newcommand{\calL}{{\cal L}}
\newcommand{\calQ}{{\cal Q}}

\newcommand{\Lcrit}{{\cal L}_{\rm crit}}
\newcommand{\Lth}{{\cal L}_{\rm th}}

\newcommand{\Rmax}{R_{\rm max}}

\newcommand{\calQp}{{\cal Q}_{\rm p}}
\newcommand{\calQe}{{\cal Q}_{\rm e}}

\newcommand{\hcalQp}{{\hat{\cal Q}}_{\rm p}}

\newcommand{\Tp}{T_{\rm p}}
\newcommand{\Te}{T_{\rm e}}

\newcommand{\hcalQe}{\hat{\cal Q}_{\rm e}}

\newcommand{\Eisco}{{\cal E}_{\rm ISCO}}

\newcommand{\Risco}{R_{\rm ISCO}}

\begin{document}

\title{Electrification of a non-rotating black hole}

\author{Ken-ichi Nakao$^{1,2,3}$\footnote{E-mail:knakao@omu.ac.jp}}
\author{Kenta Matsuo$^1$}
\author{Hirotaka Yoshino$^{1,2,3}$}
\author{Hideki Ishihara$^2$}
\affiliation{
${}^{1}$
Department of Physics, Graduate School of Science, Osaka Metropolitan University, 3-3-138 Sugimoto, Sumiyoshi, Osaka 558-8585, Japan\\
${}^{2}$Nambu Yoichiro Institute for Theoretical and Experimental Physics (NITEP), Osaka Metropolitan University, 
3-3-138 Sugimoto, Sumiyoshi, Osaka 558-8585, Japan\\
${}^{3}$Osaka Central Advanced Mathematical Institute (OCAMI), 
Osaka Metropolitan University, 3-3-138 Sugimoto, Sumiyoshi, Osaka 558-8585, Japan}


\begin{abstract}
Zaja\v{c}ek et al made an interesting theoretical prediction on the electrification of a non-rotating black hole; 
if a non-rotating black hole is surrounded by plasma composed of protons and electrons, 
it will acquire electric charge due to the large difference of the inertial mass of a proton and that of an electron.  
Furthermore they revealed the effects of the electric charge of the black hole on the surrounding plasma. 
Since their results mainly rely on non-relativistic analyses, we study the same subject through relativistic analyses in this paper. 
By investigating a test particle in the Schwarzschild spacetime, we find that if initial velocities of protons and electrons far from a black hole 
follow the Maxwell distribution, the black hole can acquire electric charge whose value depends on the ratio of temperature 
of the proton and that of the electron. 
We also show that if the black hole acquires the electric charge, the radii of the innermost stable circular orbit (ISCO) 
and the specific energy of a charged particle on ISCO can be very different from those 
of a neutral particle. In contrast to the result obtained by Zaja\v{c}ek {\it et al}, we find that the ISCO radii of a proton and an electron 
are necessarily larger than that of a neutral test particle as long as the black hole acquires the charge. The large ISCO radius 
might lead to a larger angular diameter of a black hole shadow and a different estimate of the released energy due to the accretion of plasma  
from the estimate based on the assumption of electric neutrality of the central black hole. 
\end{abstract}

\preprint{OCU-PHYS-601}
\preprint{AP-GR-199}
\preprint{NITEP 225}
\date{\today}
\maketitle

\section{Introduction}\label{sec:Intro}

The black hole is a central subject in the study of strong gravity.  It is defined as a complement of the causal past 
of the future null infinity~\cite{Hawking}, and so not observable by its definition. However, nontrivial 
physical phenomena will occur due to the strong gravitational field around it; various active astrophysical phenomena are considered to 
come from it. For example, energies of relativistic jets from active galactic nuclei (see, for example, Ref.~\cite{Brandford-etal})  
are considered to be supplied by the gravitational energy of matter accreting to the black hole \cite{Lynden-Bell,Bardeen,Rees} 
or the extraction of the rotational energy stored in the ergosphere of a rotating black hole \cite{BZ-process}. 

In the study of astrophysical black holes, they have usually been assumed to be electrically neutral. However, there are several 
studies on the significance of electrically charged black hole in the context of the astrophysics. The most  attracted electrification mechanism of a black hole 
was given by Wald \cite{Wald-solution}: if a rotating black hole with the mass $M$ and the Kerr parameter $a$, i.e., the angular momentum divided by $Mc$  
is immersed in the external homogeneous magnetic field $B$, it will be electrified until the acquired charge reaches the the value  
$8\pi\epzero GMaB/c$, where $\epzero$, $c$ and $G$ are the permittivity of vacuum, the speed of light and Newton's gravitational constant. 
Then, effects of the electric field due to the electric charge of a black hole 
have been studied in various astrophysical contexts \cite{Zajacek-etal,Levin-etal,King-Pringle,Komissarov}. 

Zaja\v{c}ek et al. studied the electrification of a non-rotating black hole \cite{Zajacek-etal}, based on mainly the non-relativistic analyses. 
Although they studied this subject also in the framework of general relativity,  
we believe that more detailed relativistic analyses of this topic are needed. 
Thus, in this paper, we explore the possibility of the electrification of a non-rotating black hole in the framework of general relativity 
by investigating  the motion of a proton and an electron as test particles in the Schwarzschild spacetime, and estimate 
the charge acquired by a non-rotating black hole in plasma. We also study the innermost stable circular orbit (ISCO) 
of the proton and the electron. 

This paper is organized as follows. In Sec.~II, we reconsider the motion of a test particle in the Schwarzschild spacetime, 
and show that if initial velocities of protons and electrons far from a black hole follow the Maxwell distribution, the probability for the proton to enter the black hole 
will be different from that of the electron, and as a result, the black hole will acquire the electric charge. 
In Sec.~III, we estimate the electric charge acquired by the black hole along the discussion on the probability made in Sec.~II. 
In Sec.~IV, the ISCO of a proton and an electron in the spacetime with the electrified black hole are shown. 
Section V is devoted to a summary. 

In this paper, we adopt the abstract index notation in which the Latin indices, except for $t$ and $r$ which are 
assigned to the time and radial coordinates, represent a type of a tensor, 
whereas the Greek indices,  except for $\theta$ and $\varphi$ which are assigned to the coordinates on a sphere, 
represent the components of a tensor with respect to the coordinate bases, 
and follow the sign convention of the metric tensor in Ref.~\cite{Wald} and the SI unit.

\section{Motion of a proton and an electron in the Schwarzschild spacetime}\label{sec:motionl}

In this section, we reconsider the motion of a test particle in the spacetime with a non-rotating black hole, i.e., 
the Schwarzschild spacetime. The infinitesimal world interval of the Schwarzschild spacetime is given as
\begin{equation}
ds^2=g_{\mu\nu}dx^\mu dx^\nu=-f(r)c^2dt^2+\frac{dr^2}{f(r)}+r^2\left(d\theta^2+\sin^2\theta d\varphi^2\right), \label{world-interval}
\end{equation}
where, with $\rg$ as a positive constant, 
\begin{equation}
f(r)=1-\frac{\rg}{r}. \label{Sch-f}
\end{equation}
The constant $\rg$ is called the gravitational radius, or the Schwarzschild radius, which is related to the gravitational mass of the system $M$ through 
\begin{equation}
\rg=\dfrac{2GM}{c^2},
\end{equation}
and is the radius of the black hole which corresponds to the domain of $r\leq\rg$. 

The world line of a massive test particle is a timelike geodesic whose unit tangent is called the four-velocity of the particle and denoted 
by $u^\mu=\dfrac{dx^\mu}{d\tau}$, where $x^\mu$ and $\tau$ represent the coordinates and the proper time of the particle, respectively.  
The four-velocity is determined by the geodesic equation
\begin{equation}
u^a\nabla_a u_b=0, \label{geodesic-eq}
\end{equation}
where $\nabla_a$ is the covariant derivative.  

Since the time and azimuthal coordinate bases are Killing vectors, the time and azimuthal components of Eq.~\eqref{geodesic-eq} are
\begin{equation}
\frac{d u_t}{d\tau}=0~~~~{\rm and}~~~~\frac{d u_\varphi}{d\tau}=0,
\end{equation}
respectively. Thus, with $E$ and $L$ as constants, we have
\begin{equation}
cf\frac{dt}{d\tau}=E~~~{\rm and}~~~r^2\sin^2\theta\frac{d\varphi}{d\tau}=L_z.
\end{equation}
Substituting these results into the $\theta$-component of the geodesic equations, we have
\begin{equation}
\frac{d}{d\tau}\left[\left(r^2\frac{d\theta}{d\tau}\right)^2+\frac{L_z^2}{\sin^2\theta}\right]=0.
\end{equation}
An integration of this equation leads to
\begin{equation}
r^2\left(\frac{d\theta}{d\tau}\right)^2+\frac{L_z^2}{r^2\sin^2\theta}=\frac{L^2}{r^2},
\end{equation}
where $L$ is a constant of integration. 
The integration constants $E$, $L_z$ and $L$ correspond to the specific energy divided by the speed of light $c$,  
$z$-component of the specific angular momentum and the total specific angular momentum, respectively. 

Here, for notational simplicity, we introduce following dimensionless quantities. 
\begin{align}
R&:=\frac{r}{\rg}, \label{dimless-r}\\
\calT&:=\frac{c\tau}{\rg}, \label{dimless-tau}\\
\calE&:=\frac{E}{c}, \label{dimless-E}\\
\calL&:=\frac{L}{c\rg}. \label{dimless-L}
\end{align}
Then the  normalization condition of the 
four-velocity $g_{\mu\nu}u^\mu u^\nu=-c^2$ leads to
\begin{equation}
\left(\frac{dR}{d\calT}\right)^2+V(R;\calE,\calL^2)=0, \label{energy-eq-n}
\end{equation}
where
\begin{align}
V(R;\calE,\calL^2)&=-\calE^2+\left(1-\frac{1}{R}\right)\left(1+\frac{\calL^2}{R^2}\right) \nonumber \\
&=1-\calE^2-\frac{1}{R}+\frac{\calL^2}{R^2}-\frac{\calL^2}{R^3} \label{V-definition}
\end{align}
is the effective potential. The allowed domain for the motion of the test particle is determined by an inequality $V(R;\calE,\calL^2)=-(dR/d\tau)^2\leq0$. 

In this section, we consider the situation in which the test particle falls from infinity. 
Thus, $\calE$ has to be larger than or equal to one. Such motion is called the unbound or marginally bound one. 
Here, note that $\calE$ has to be positive so that $t$ is an increasing function of the proper time $\tau$ 
in the domain outside the black hole, $R>1$. 
The extremum of the effective potential is determined by $\partial_RV(R;\calE,\calL)=0$ which is equivalent to 
\begin{equation}
R^2-2\calL^2 R+3\calL^2=0. \label{V-extrema-eq}
\end{equation}
The roots of this equation are given as
\begin{equation}
R=R_\pm:=\calL^2\left(1\pm\sqrt{1-\frac{3}{\calL^2}}\right). \label{V-extrema}
\end{equation}
We can easily see that the derivative of the effective potential with respect to $R$ is positive for both $R\rightarrow 0$ and $R\rightarrow\infty$. 
In the case of $\calL^2\leq3$, since $R_\pm$ are complex, there is no extremum, and hence, the test particle enters the black hole. 
On the other hand, if $\calL^2>3$ holds, $V(R;\calE,\calL)$ takes the maximum and the minimum at $R=R_-$ and $R=R_+$, respectively. 
In this case, the test particle enters the black hole if and only if $V(R_-;\calE,\calL)<0$ holds, 
or equivalently,  
\begin{equation}
\calE^2-1>\lambda(\calL^2), \label{IN-condition}
\end{equation}
where
\begin{align}
\lambda(\calL^2)&:=-\frac{1}{R_-}+\frac{\calL^2}{R_-^2}-\frac{\calL^2}{R_-^3} 
=\frac{\calL^2}{27}\left[2-\frac{9}{\calL^2}+2\left(1-\frac{3}{\calL^2}\right)^{3\over2}\right]. \label{lambda-def}
\end{align}

We consider the situation in which the velocity of the test particle is much less than the speed of light at $r\gg\rg$ ($R\gg1$) 
and follows  the Maxwell distribution with the temperature $T$; 
\begin{equation}
f(\bm{v})=\left(\frac{m}{2\pi \kB T}\right)^{3\over2}\exp\left(-\frac{m} {2\kB T}\bm{v}^2\right), \label{Maxwell}
\end{equation}
where $m$ is the mass of the test particle, and the three-velocity $\bm{v}$ is defined as the spatial components of the four-velocity $u^a$ with respect to 
the orthonormal frame of the static observer (see Appendix \ref{orthonormal} for more details);
\begin{align}
v_{(r)}&=\frac{1}{\sqrt{f(r)}}\frac{dr}{d\tau}, \label{vr-def}\\
v_{(\theta)}&=r \frac{d\theta}{d\tau}, \\
v_{(\varphi)}&=r\sin\theta\frac{d\varphi}{d\tau}. \label{vp-def}
\end{align}
We have
\begin{align}
\bm{v}^2&=v_{(r)}^2+v_{(\theta)}^2+v_{(\varphi)}^2 \nonumber\\
&=\frac{1}{f(r)}\left(\frac{dr}{d\tau}\right)^2+r^2\left(\frac{d\theta}{d\tau}\right)^2+r^2\sin^2\theta\left(\frac{d\varphi}{d\tau}\right)^2 \nonumber \\
&=\frac{1}{f(r)}\left(\frac{dr}{d\tau}\right)^2+\frac{L^2}{r^2} \nonumber\\
&=\frac{c^2}{f(r)}\left[\calE^2-f(r)\right]
\end{align}
and so
\begin{equation}
\calE^2-1=f(r)\dfrac{\bm{v}^2}{c^2}-\dfrac{\rg}{r}.
\label{E-square}
\end{equation}
Then, Eq.~\eqref{IN-condition} leads to
\begin{equation}
\frac{v_{(r)}^2}{c^2}>\left(1-\frac{1}{R}\right)^{-1}\left[\lambda\left(\calL^2\right)+\frac{1}{R}\right]-\frac{\calL^2}{R^2}.
\end{equation}
For $R\gg1$, we have 
\begin{equation}
\frac{v_{(r)}^2}{c^2}>\lambda\left(\calL^2\right)+{\cal O}\left(R^{-1}\right), \label{vr-domain}
\end{equation}
or equivalently,
\begin{equation}
\calL^2<\lambda^{-1}\left(\frac{v_{(r)}^2}{c^2}\right)+{\cal O}\left(R^{-1}\right), \label{L-domain}
\end{equation}
where $x=\lambda^{-1}(y)$ is the inverse function of $y=\lambda(x)$. 

We introduce $\Lcrit^2$ as a positive root of the following equation,
\begin{equation}
\lambda\left(\Lcrit^2\right)=0, \label{Lcrit-n}
\end{equation}
We have the root of this equation as
\begin{equation}
\Lcrit^2=4 \label{Lcrit-sol-n}
\end{equation}
In the case of $3\leq\calL^2\leq\Lcrit^2$, since $\lambda(\calL^2)\leq0$ holds, Eq.~\eqref{vr-domain} is trivial.  
Since the allowed domain for the motion of the charged test particle with $\calL^2<3$ should be $0<R<\infty$,  
the test particle enters the black hole if the initial radial velocity is less than or equal to zero in the case of $0\leq\calL^2\leq\Lcrit^2$.  
Only in the case of $\calL^2>\Lcrit^2$, Eq.~\eqref{vr-domain} gives a non-trivial condition.

Here, we estimate the probability $P$ for the particle to enter the black hole. Instead of $v_{(\theta)}$ and $v_{(\varphi)}$, 
we adopt the variables defined as
\begin{align}
v_{(\theta)}&=\vomg\cos\Theta, \\
v_{(\varphi)}&=\vomg\sin\Theta,
\end{align}
where $0\leq \vomg<\infty$ and $0\leq\Theta<2\pi$. Note that 
\begin{equation}
\frac{\vomg^2}{c^2}=\frac{\calL^2}{R^2}. \label{V-L-relation}
\end{equation}

The infinitesimal volume element in the velocity space is given as
\begin{equation}
dv_{(r)}dv_{(\theta)}dv_{(\varphi)}=dv_{(r)}\vomg d\vomg d\Theta = \frac{c^2}{2R^2}dv_{(r)} d\calL^2 d\Theta,
\end{equation}
where Eq.~\eqref{V-L-relation} has been used in the last equality. 
The square of the dimensionless angular momentum $\calL^2$ should satisfy Eq.~\eqref{L-domain} and $v_{(r)}$ should be 
less than or equal to zero so that the  test particle enters the black hole. Then, the probability for the test particle to enter the black hole is given as
\begin{align}
P=&\int_0^{2\pi}d\Theta\int_{-\infty}^0dv_{(r)}\left(\frac{m}{2\pi \kB T}\right)^{1\over2}\exp\left(-\frac{m} {2\kB T}v_{(r)}^2\right) \nonumber \\
&~~~~~~~~~~~~~~~~~~~~~~~~~~~~~~~~~\times
\int_0^{\lambda^{-1}\left(v_{(r)}^2/c^2\right)}d\calL^2 \frac{mc^2}{4\pi \kB T R^2}\exp\left(-\frac{mc^2} {2\kB T R^2}\calL^2\right) \nonumber \\
=&
\int_{-\infty}^0dv_{(r)}\left(\frac{m}{2\pi \kB T}\right)^{1\over2}\exp\left(-\frac{m} {2\kB T}v_{(r)}^2\right) 
\left\{1-\exp\left[-\frac{mc^2} {2\kB T R^2}\lambda^{-1}\left(\frac{v_{(r)}^2}{c^2}\right)\right]\right\} \nonumber \\
=&\int_0^\infty dv_{(r)}\left(\frac{m}{2\pi \kB T}\right)^{1\over2}\exp\left(-\frac{m} {2\kB T}v_{(r)}^2\right) 
\left[\frac{mc^2} {2\kB T R^2}\lambda^{-1}\left(\frac{v_{(r)}^2}{c^2}\right)+{\cal O}\left(R^{-4}\right) \right],
\label{Prob-n}
\end{align}
where, in the final equality, $|v_{(r)}/c|$ has been assumed to be much less than unity, since 
we assume $mc^2/\kB T\gg1$. 

By solving Eq.~\eqref{lambda-def} with respect to $\calL^2$, we have
\begin{equation}
\lambda^{-1}(x)=\frac{1}{8x}\left(27x^2+18x-1+\sqrt{\left(27x^2+18x-1\right)^2+64x}\right).
\end{equation}
Hence, we have
\begin{equation}
\lambda^{-1}\left(\frac{v_{(r)}^2}{c^2}\right)=4+\frac{8v_{(r)}^2}{c^2}+{\cal O}\left(\frac{v_{(r)}^4}{c^4}\right)
\end{equation}
and 
\begin{align}
P&=\frac{2mc^2}{\kB T R^2}\times\frac{1}{\sqrt{\pi}}\int_0^\infty\left[1+\frac{4\kB T}{mc^2}x^2
+{\cal O}\left(\left(\frac{\kB T}{mc^2}\right)^2\right)\right] e^{-x^2}dx \nonumber \\
&=\frac{mc^2}{\kB T R^2}\left[1
+{\cal O}\left(\frac{\kB T}{mc^2}\right)\right].
\label{P-neutral}
\end{align}

Hereafter the test particle is assumed to be a proton or an electron. 
Any quantity of a proton is denoted by the character with a subscript ``p", whereas 
that of an electron is denoted by the character with a subscript ``e". 
Thus the excess of the probability of the proton to enter the black hole is given as
\begin{equation}
\Delta P=P_{\rm p}-P_{\rm e}\simeq \frac{1}{R^2}\left(\frac{\mpr c^2}{\kB\Tp}-\frac{\mel c^2}{\kB\Te}\right).
\label{Number-n}
\end{equation}

If $\mpr/\Tp >\mel /\Te$ holds, $\Delta P$ is positive and the black hole will be positively electrified.  
By contrast, if $\mpr/\Tp < \mel/\Te$ holds, $\Delta P$ is negative and the black hole will be negatively electrified. 
If $\mpr/\Tp =\mel/\Te$ holds, no electrification of the black hole will occur. 
If the temperature of the protons and that of the electrons are identical to each other,  the black hole will be positively electrified 
because of $\mpr\gg\mel$. 
 
Let us consider the Coulomb force between protons and electrons, which is ignored in our consideration. 
The mean separation between protons and electrons is denoted by $l$. 
Then, the relative acceleration $a_{\rm C}$ between a proton and a neighboring electron due to Coulomb force is estimated as
\begin{equation}
a_{\rm C}=\frac{e^2}{4\pi\epzero l^2}\frac{\mel+\mpr}{\mel\mpr} \simeq\frac{e^2}{4\pi\epzero \mel l^2},
\end{equation}
whereas, for the proton and the neighboring electron on a sphere of $r=L$, the relative acceleration $a_{\rm T}$ due to the tidal force is given as
\begin{equation}
a_{\rm T}
\simeq \frac{2GMl}{L^3},
\end{equation}
where we have assumed $l\ll L$. 
Our present consideration is justified only if $a_{\rm C}<a_{\rm T}$, or equivalently, 
\begin{equation}
l>\left(\frac{\alpha\hbar \rg^2}{\mel c}\right)^{1\over3}\frac{L}{\rg}=7.3\times10^2 \left(\frac{L}{10\rg}\right)\left(\frac{M}{4\times10^6M_\odot}\right)^{2\over3}{\rm m}
\end{equation}
holds, where $\hbar$ and $\alpha$ are the Dirac constant and the fine structure constant $e^2/4\pi\epzero\hbar c\simeq1/137$, respectively. 

The spatial curvature of our Universe is almost equal to zero \cite{Planck2018}, and thus, from the observed baryon to photon ratio, 
the average density of baryonic matter is estimated at 
$\Omega_{\rm b}=0.0224h^{-2}$ times the critical density $3H_0^2/8\pi G$, where $H_0=100 h$(km/s/Mpc) 
is the Hubble constant.\footnote{Note that $h$ is not the Planck constant but is equal to $67.66\pm0.42$ \cite{Planck2018}.} 
Assuming that the energy density of baryonic matter is almost equal to that of protons, the mean separation of protons is equal to about 1.6m, and hence 
our result without any modification cannot be applied to the black hole with the mass comparable to that of Sgr A* for which we need to take 
into account the electromagnetic interaction between protons and electrons. However, since the lower bound on $l$ is about $7.3$ cm in the case of 
the black hole of $4M_\odot$, there may be situations for stellar mass black holes 
in our Universe to which our consideration is applicable without any modification.

\section{Maximal charge acquired by non-rotating black hole}\label{sec:max-charge}

In this section, we estimate the charge acquired by the black hole in the situation assumed in the previous section. 
The selective accretion of protons or electrons to the black hole will cease by the electromagnetic repulsion after the black hole 
acquires sufficient electric charge. 

The non-rotating charged black hole is described by the Reissner-Nordstr\"{o}m solution 
whose infinitesimal world interval is also given by Eq.~\eqref{world-interval} but with 
\begin{equation}
f(r)=1-\frac{\rg}{r}+\frac{Q^2}{r^2}, \label{RN-f}
\end{equation}
where $Q$ is the charge parameter which is related to the electric charge $\qE$ of the black hole through 
\begin{equation}
Q^2=\frac{G\qE^2}{4\pi\epzero c^4}. 
\end{equation}
As will be shown later, the maximal charge acquired by the black hole is so small that 
$\rg \gg \dfrac{Q^2}{r}$ holds for $r>\rg$. Hence, we need not use Eq.~\eqref{RN-f} 
but Eq.~\eqref{Sch-f} in the following analyses. 
By contrast, the electric field produced by $\qE$ can strongly affect the motion of a proton and an electron. 
The four-vector potential $A_a$ is given as
\begin{equation}
A_a=-\frac{\qE}{4\pi\epzero c r} \delta^t_a,
\end{equation}
where $\delta_a^b$ is the Kronecker delta. 

The world line of a charged test particle is not a geodesic in the situation we consider. The equation of motion is  
\begin{equation}
m u^a\nabla_a u_b=q F_{bc}u^c, \label{EOM}
\end{equation}
where $m$ and $q$ are the mass and the electric charge of the test particle, and, 
with $\partial_b$ as the ordinary partial derivative, $F_{bc}=\partial_b A_c-\partial_c A_b$ is 
the electromagnetic field strength tensor. 
Since the time and the azimuthal coordinate bases are Killing vectors, 
the time and azimuthal components of Eq.~\eqref{EOM} are
\begin{align}
\frac{d}{d\tau}\left(u_t +\frac{q}{m} A_t\right)=0~~~~~{\rm and}~~~~~\frac{d u_\varphi}{d\tau}=0. 
\end{align}
Thus, as in the previous section, we have, with $E$ and $L_z$ as constants of integration, we have
\begin{equation}
-u_t-\frac{q}{m}A_t=E~~~{\rm and}~~~u_\varphi=L_z.
\end{equation}
Substituting these results into the $\theta$-component of Eq.~\eqref{EOM}, we have
\begin{equation}
\frac{d}{d\tau}\left[\left(r^2\frac{d\theta}{d\tau}\right)^2+\frac{L_z^2}{\sin^2\theta}\right]=0.
\end{equation}
An integration of this equation leads to
\begin{equation}
r^2\left(\frac{d\theta}{d\tau}\right)^2+\frac{L_z^2}{r^2\sin^2\theta}=\frac{L^2}{r^2},
\end{equation}
where $L$ is a constant of integration. 
The constants of integration, $E$, $L_z$ and $L$, are the specific energy devided by the speed of light $c$, 
$z$-component of the specific angular momentum and the total specific angular momentum, respectively. 

As in the previous section, we adopt the dimensionless quantities defined as Eqs.~\eqref{dimless-r}--\eqref{dimless-L}. 
We introduce one more dimensionless quantity related to 
the electric charges of the black hole and the test particle, which is defined as
\begin{equation}
\calQ:=\frac{q\qE}{4\pi\epzero m c^2\rg}.
\end{equation}
Since we have already studied the case of $\calQ=0$, we will consider the case of $\calQ\neq0$ in this section. 

Then, the normalization condition of the four-velocity $g_{\mu\nu}u^\mu u^\nu=-c^2$ leads to the similar equation to Eq.~\eqref{energy-eq-n} as
\begin{equation}
\left(\frac{dR}{d\calT}\right)^2+U(R;\calE,\calL^2,\calQ)=0, \label{energy-eq-c}
\end{equation}
where
\begin{align}
U(R;\calE,\calL^2,\calQ)
&=-\left(\calE-\frac{\calQ}{R}\right)^2+\left(1-\frac{1}{R}\right)\left(1+\frac{\calL^2}{R^2}\right) 
\label{potential-c}
\end{align}
is the effective potential for the charged particle.   
The allowed domain for the motion of the charged test particle is determined by an inequality $U(R;\calE,\calL^2,\calQ)=-(dR/d\calT)^2\leq0$. 

Here note that since $u_t=-f(r)d(ct)/d\tau$ should be negative outside the black hole, $\calE-\calQ/R$ should be positive in the allowed 
domain for the motion of the test particle. This fact implies that $R>\calQ/\calE$ must hold in the allowed domain. 
In order that the infinity $R\rightarrow\infty$ is included in the allowed domain, both $\calE$ and $\calE^2-1$ 
must be positive. Thus, $\calE\geq1$ should hold in the situation of our interest. 

We rewrite the effective potential $U$ in the form,
\begin{equation}
U(R;\calE,\calL^2,\calQ)=\Bigl(G_-(R;\calL^2,\calQ)+\calE\Bigr)\Bigl(G_+(R;\calL^2,\calQ)-\calE\Bigr),
\end{equation}
where
\begin{equation}
G_\pm(R;\calL^2,\calQ)=\sqrt{\left(1-\frac{1}{R}\right)\left(1+\frac{\calL^2}{R^2}\right)}~\pm\frac{\calQ}{R}.
\end{equation}
Note that $G_\pm$ is defined in the domain $R\geq1$. Since, as mentioned, $\calE-\calQ/R$ should be positive, 
$G_-+\calE$ is positive. Hence the necessary condition $U(R;\calE,\calL^2,\calQ)\leq 0$ 
for the domain allowed for the motion of the charged test particle is equivalent to 
\begin{equation}
G_+(R;\calL^2,\calQ) \leq \calE.  \label{condition-0}
\end{equation}

By investigating numerically and analytically the behavior of $G_+$, we find the following facts.
\begin{enumerate}
 \item In the case of $\calQ\leq\dfrac{1}{2}$, $G_+$ has a maximum as a function of $R$ 
 if and only if $\calL^2$ is larger than a threshold $\Lth^2$. On the other hand, if $\calL^2$ is less than or equal to $\Lth^2$, 
 $G_+$ is a non-decreasing function of $R$. 
 \item In the case of $\calQ=\dfrac{1}{2}$, the threshold $\Lth^2$ is equal to $\dfrac{1}{4}$. 
 \item In the case of $\calQ>\dfrac{1}{2}$, $G_+$ has a maximum as a function of $R$ for any $\calL^2$. The maximum value of $G_+$ is positive.  
 \end{enumerate}

If $G_+$ has a maximum, the dimensionless radius $R$ at which $G_+$ takes 
the maximum value is denoted by $\Rmax(\calL^2;\calQ)$. Then we have
\begin{align}
\frac{\partial G_+\bigl(\Rmax\left(\calL^2;\calQ\right);\calL^2,\calQ\bigr)}{\partial \calL^2}
&=\left.\left(\frac{\partial \Rmax\left(\calL^2;\calQ\right)}{\partial \calL^2}\frac{\partial G_+\left(R;\calL^2,\calQ\right)}{\partial R}
+\frac{\partial G_+\left(R;\calL^2,\calQ\right)}{\partial \calL^2}\right)\right|_{R=\Rmax} \nonumber \\
&=\frac{1}{2\Rmax^2}\sqrt{\frac{\Rmax(\Rmax-1)}{\Rmax^2+\calL^2}}>0, 
\label{delG-delL}
\end{align}
where we have used $\partial G_+/\partial R|_{R=\Rmax}=0$ in the second equality. 
 Thus, when there is a maximum of $G_+$ as a function of $R$, the maximum value of $G_+$ is an increasing function of $\calL^2$. 

We define a function $\Lambda(\calL^2;\calQ)$ as
\begin{equation}
\Lambda(\calL^2;\calQ):=G_+^2\Bigl(\Rmax(\calL^2;\calQ);\calL^2,\calQ\Bigr)-1. \label{Lambda-def}
\end{equation}
We can easily see from Eq.~\eqref{delG-delL} that $\Lambda(\calL;\calQ)$ is an increasing function of $\calL^2$ 
if and only if the maximum value of $G_+$ is non-negative.  In Appendix \ref{G-plus}, we show
\begin{equation}
\lim_{\calL^2\rightarrow\infty}\Rmax(\calL^2,\calQ)=\frac{3}{2}
\end{equation}
and
\begin{equation}
\Lambda(\calL^2,\calQ)\longrightarrow \frac{4}{27}\calL^2~~~{\rm for}~~~\calL^2\gg1. \label{Lambda-asymptotic} 
\end{equation}

Note that $\Lambda\left(\calL^2;\calQ\right)$ is a corresponding quantity to $\lambda\left(\calL^2\right)$ introduced in the case of the 
electrically neutral test particle. If $G_+$ has a positive maximum, 
the necessary and sufficient condition for the charged test particle to enter the black hole is
\begin{equation}
\Lambda(\calL^2;\calQ) < \calE^2-1. \label{Lambda-condition}
\end{equation}

By using the three-velocity defined as Eqs.~\eqref{vr-def}--\eqref{vp-def}, the normalization of the four-velocity leads to 
\begin{equation}
\calE-\frac{Q}{R}=\sqrt{\left(1-\frac{1}{R}\right)\left(1+\frac{\calL^2}{R^2}+\frac{v_{(r)}^2}{c^2}\right)},
\end{equation}
where we have used the fact that $\calE-Q/R$ should be positive. Hence, Eq.~\eqref{Lambda-condition} is rewritten in the form
\begin{equation}
\Lambda\left(\calL^2;Q\right) < \left[\sqrt{\left(1-\frac{1}{R}\right)\left(1+\frac{\calL^2}{R^2}+\frac{v_{(r)}^2}{c^2}\right)}+\frac{Q}{R}\right]^2-1.
\end{equation}
Since we focus on the case of $R\gg1$, by virtue of Eq.~\eqref{Lambda-asymptotic}, we have 
\begin{equation}
\frac{v_{(r)}^2}{c^2}>\Lambda\left(\calL^2;Q\right)+{\cal O}\left(R^{-1}\right), \label{vr-domain-c}
\end{equation}
or equivalently,
\begin{equation}
\calL^2<\Lambda^{-1}\left(\frac{v_{(r)}^2}{c^2};\calQ\right)+{\cal O}\left(R^{-1}\right), \label{L-domain-c}
\end{equation}
where $x=\Lambda^{-1}(y;\calQ)$ is the inverse function of $y=\Lambda(x;\calQ)$. 

We consider the case of $\calQ\leq1/2$ and that of $\calQ>1/2$ separately below.

\subsection{The case of $\calQ\leq1/2$}\label{Q-le-half}

We introduce a quantity $\Lcrit(\calQ)$ as a positive root of the following equation,
\begin{equation}
G_+\Bigl(\Rmax(\Lcrit^2;\calQ);\Lcrit^2,\calQ\Bigr)-1=0, 
\end{equation}
or equivalently, 
\begin{equation}
\sqrt{\left(1-\frac{1}{\Rmax}\right)\left(1+\frac{\calL^2}{\Rmax^2}\right)}=1-\frac{\calQ}{\Rmax}. \label{0th-Lcrit}
\end{equation} 
Taking a square of both sides of this equation, we obtain a quadratic equation for $\Rmax$ as
\begin{equation}
(1-2\calQ)\Rmax^2-\left(\Lcrit^2-\calQ^2\right)\Rmax+\Lcrit^2=0. \label{1st-Lcrit}
\end{equation}
Since $\Rmax$ should be a degenerate root of this quadratic equation, the discriminant should vanish;
\begin{equation}
(\Lcrit^2-\calQ^2)^2-4(1-2\calQ)\Lcrit^2=0.
\end{equation}
The roots of this equation for $\calL^2$ are
\begin{equation}
\Lcrit^2=\calL_\pm^2:=\calQ^2-4\calQ+2\pm2\sqrt{(1-\calQ)^2(1-2\calQ)}. \label{Lcrit-roots}
\end{equation}
Note that 
\begin{align}
\calL_-^2-\calQ^2&=2-4\calQ-2\sqrt{(1-\calQ)^2(1-2\calQ)} \nonumber \\
&=-\frac{2\calQ^2(1-2\calQ)}{1-2\calQ+\sqrt{(1-\calQ)^2(1-2\calQ)}} \nonumber \\
&\leq0  \label{sign}
\end{align}
holds, whereas $\calL_+^2-\calQ^2>0$ is trivial.  
The degenerate root of Eq.~\eqref{1st-Lcrit} is equal to either $\left(\calL_+^2-\calQ^2\right)/2(1-2\calQ)$ or$\left(\calL_-^2-\calQ^2\right)/2(1-2\calQ)$. 
By Eq.~\eqref{sign}, we find that the second one is negative, but $\Rmax$ should be positive. Thus we have
\begin{equation}
\Lcrit^2(\calQ)=\calL_+^2 \label{Lcrit-sol-c}
\end{equation}
and
\begin{equation}
\Rmax\left(\Lcrit^2;\calQ\right)=\frac{\calL_+^2-\calQ^2}{2(1-2\calQ)}. \label{Rmax-crit}
\end{equation}

Here note that 
\begin{align}
\Lambda\left(\Lcrit^2;\calQ\right)&=\left[G_+\Bigl(\Rmax(\Lcrit^2,\calQ);\Lcrit^2,\calQ\Bigr)+1\right]\left[G_+\Bigl(\Rmax(\Lcrit^2,\calQ);\Lcrit^2,\calQ\Bigr)-1\right] \nonumber \\
&=0
\end{align}
holds. 
Due to Eq.~\eqref{delG-delL}, $\Lambda$ is an increasing function of $\calL^2$ 
as long as $G_+$ is positive. Hence $\Lambda$ is non-negative for $\calL^2\geq\Lcrit^2$, while it is negative for $\calL^2<\Lcrit^2$ and so  
the inequality \eqref{vr-domain-c} gives no constraint on $v_{(r)}^2$ for $\calL^2<\Lcrit^2$.  
By contrast, Eq.~\eqref{L-domain-c} gives a constraint on $\calL^2$ for any $v_{(r)}^2$. 

As we derived Eq.~\eqref{Prob-n} in the case of the neutral test particle,  
the probability $P$ for the charged test particle of $\calQ\leq1/2$ to enter the black hole is given as
\begin{align}
P=&\int_0^{2\pi}d\Theta\int_{-\infty}^0dv_{(r)}\left(\frac{m}{2\pi \kB T}\right)^{1\over2}\exp\left(-\frac{m} {2\kB T}v_{(r)}^2\right) \nonumber \\
&~~~~~~~~~~~~~~~~~~~~~~~~~~~~~~~\times
\int_0^{\Lambda^{-1}\left(v_{(r)}^2/c^2;\calQ\right)}d\calL^2 \frac{mc^2}{4\pi \kB T R^2}\exp\left(-\frac{mc^2} {2\kB T R^2}\calL^2\right) \nonumber \\
=&
\int_{-\infty}^0dv_{(r)}\left(\frac{m}{2\pi \kB T}\right)^{1\over2}\exp\left(-\frac{m} {2\kB T}v_{(r)}^2\right) 
\left\{1-\exp\left[-\frac{mc^2} {2\kB T R^2}\Lambda^{-1}\left(\frac{v_{(r)}^2}{c^2};\calQ\right)\right]\right\} \nonumber \\
=&\int_0^\infty dv_{(r)}\left(\frac{m}{2\pi \kB T}\right)^{1\over2}\exp\left(-\frac{m} {2\kB T}v_{(r)}^2\right) 
\left[\frac{mc^2} {2\kB T R^2}\Lambda^{-1}\left(\frac{v_{(r)}^2}{c^2};\calQ\right)+{\cal O}\left(R^{-4}\right) \right],
\label{Prob-c}
\end{align}
where we have assumed $R\gg\sqrt{mc^2/\kB T}$ in the last equality. 
Since we focus on the non-relativistic situation $mc^2/\kB T\gg1$, $v_{(r)}^2/c^2$ can be assumed to be 
much smaller than unity in the integrand of the last equality 
in Eq.~\eqref{Prob-c}. Thus, we have
\begin{align}
P\simeq&\int_0^\infty dv_{(r)}\left(\frac{m}{2\pi \kB T}\right)^{1\over2}\exp\left(-\frac{m} {2\kB T}v_{(r)}^2\right) 
\frac{mc^2}{2\kB T R^2}\Lcrit^2=\frac{\calL_+^2}{4R^2}\frac{mc^2}{\kB T}. \label{P-charged}
\end{align}

We consider the difference between the probabilities of the proton and the electron to enter the black hole. 
For notational simplicity, we introduce the following small dimensionless constant
\begin{equation}
\epsilon=\frac{\mel}{\mpr}=5.446\times10^{-4}.
\end{equation}
We have
\begin{equation}
\calQp=\frac{e\qE}{4\pi\epzero \mpr c^2\rg}= \epsilon\times\frac{e\qE}{4\pi\epzero \mel c^2\rg}=-\epsilon\calQe.
\end{equation}
Then, from Eq.~\eqref{Lcrit-roots}, we obtain
\begin{align}
\calL^2_{\rm +p}&=\calQp^2-4\calQp+2+2\sqrt{\left(1-\calQp\right)^2\left(1-2\calQp\right)}, \label{calLp} \\
\calL^2_{\rm +e}&=\calQe^2-4\calQe+2+2\sqrt{\left(1-\calQe\right)^2\left(1-2\calQe\right)}\nonumber \\
&=\epsilon^{-2}\left[\calQp^2+4\epsilon\calQp+2\epsilon^2
+2\epsilon^{1\over2}\sqrt{\left(\epsilon+\calQp\right)^2\left(\epsilon+2\calQp\right)}\right]. \label{calLe}
\end{align}
Furthermore, we introduce a parameter defined as
\begin{equation}
\kappa:=\frac{\Te}{\Tp}.
\end{equation}

Then, as Eq.~\eqref{Number-n}, we can estimate the excess of the probability of the proton to enter the black hole at 
\begin{equation}
\Delta P=\Pp-\Pe
\simeq\frac{1}{4R^2}\left(\calL_{\rm +p}^2\frac{\mpr c^2}{\kB\Tp}-\calL_{\rm +e}^2\frac{\mel c^2}{\kB\Te}\right)
~\left\{
\begin{array}{lll}
>0 &{\rm for}~\kappa>\epsilon\calL_{\rm +e}^2/\calL_{\rm +p}^2 \\
=0 &{\rm for}~\kappa=\epsilon\calL_{\rm +e}^2/\calL_{\rm +p}^2 \\
<0 &{\rm for}~\kappa<\epsilon\calL_{\rm +e}^2/\calL_{\rm +p}^2
\end{array}
\right. .
\label{Number-c}
\end{equation}
Hence the electrification of the black hole will cease, when $\calQp$ becomes $\hcalQp$ which is a root of the following equation
\begin{equation}
\kappa=\epsilon \frac{\calL_{\rm +e}^2}{\calL_{\rm +p}^2}
=\epsilon^{-1}\frac{\hcalQp^2+4\epsilon\hcalQp+2\epsilon^2+2\epsilon^{1\over2}\sqrt{\left(\epsilon+\hcalQp\right)^2\left(\epsilon+2\hcalQp\right)}}
{\hcalQp^2-4\hcalQp+2+2\sqrt{\left(1-\hcalQp\right)^2\left(1-2\hcalQp\right)}}.
 \label{condition-c}
\end{equation}
Equation \eqref{condition-c} determines the electric charge which will be acquired by the non-rotating black hole 
in the assumed situation (see Appendix \ref{End} for more detail).

\begin{figure}[!h]
\centering\includegraphics[width=9cm]{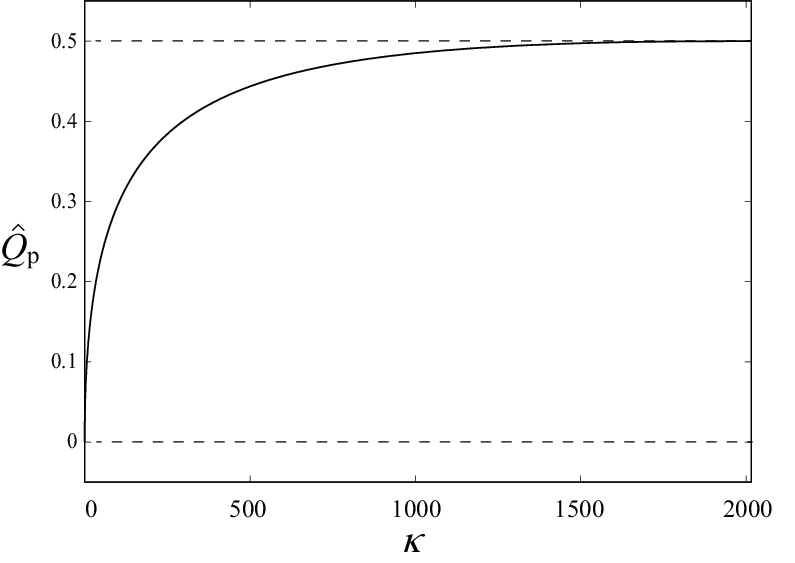}
\caption{ The dimensionless charge parameter of the proton $\hcalQp$ at which the selective accretion will cease 
is depicted as a function of $\kappa$ for the domain Eq.~\eqref{kappa-domain}. 
}
\label{fig:kappa-charge}
\end{figure}

We consider the case of $\calQp\leq 1/2$ for the proton and $\calQe\leq1/2$ for the electron. 
Because of $\calQe=-\epsilon^{-1}\calQp$, $\calQe\leq1/2$ leads to $\calQp\geq-\epsilon/2$. Hence we have
\begin{equation}
-\frac{\epsilon}{2}\leq \calQp \leq \frac{1}{2}. \label{condition-5}
\end{equation} 

In the situation assumed here, 
\begin{equation}
q_{\rm min}\leq \qE \leq q_{\rm max}
\end{equation}
holds, where
\begin{align}
\qmin&=-\frac{4\pi\epzero GM\mel }{e}=-3.36\times10^5\left(\frac{M}{4\times10^6M_\odot}\right){\rm C}.\label{minimal-c}\\
\qmax&=\frac{4\pi\epzero GM\mpr }{e}=6.17\times10^8\left(\frac{M}{4\times10^6M_\odot}\right){\rm C}. \label{maximal-c}
\end{align}
The charge parameter of the Reissner-Nordstr\"{o}m solution, $Q$, corresponding to $\qmax$ is given as
\begin{equation}
Q^2=\frac{1}{4\alpha}\left(\frac{\mpr}{m_{\rm pl}}\right)^2\rg^2=2.0\times10^{-37}\rg^2.
\end{equation}
This result justifies the neglect of the charge term $Q^2/r^2$ in $f(r)$ of Eq.~\eqref{RN-f}.

We depict $\hcalQp$ as a function of $\kappa$ in Fig.~\ref{fig:kappa-charge}. 
Since $\hcalQp$ is an increasing function of $\kappa$, Eq.~\eqref{condition-5} leads to an inequality for $\kappa$ as
\begin{equation}
\kappa_{\rm min}\leq\kappa\leq\kappa_{\rm max}, \label{kappa-domain}
\end{equation}
where
\begin{align}
\kappa_{\rm min}&=\epsilon\left[8+4\sqrt{(2+\epsilon)^2(1+\epsilon)}+8\epsilon+\epsilon^2\right]^{-1}, \\
\kappa_{\rm max}&=\epsilon^{-1}+4\epsilon^{-1/2}\sqrt{(1+2\epsilon)^2(1+\epsilon)}+8+8\epsilon.
\end{align}
Figure 2 is basically the same as Fig.~1 but, for clarity, we depict $\epsilon^{-1} \hcalQp=-\hcalQe$ 
in the domain $\kappa_{\rm min}\leq \kappa<10^{-3}$. The dimensionless charge parameter $\hcalQp$  
is negative for $\kappa<\epsilon$, vanishes at $\kappa=\epsilon$ and is positive for $\kappa>\epsilon$. 
The selective accretion of protons or electrons may occur except in the special case of $\Te=\epsilon\Tp\ll\Tp$. 

Note that if $\kappa<\kappa_{\rm min}$ holds, there is no root of Eq.~\eqref{condition-c} for $\hcalQp$ in the domain \eqref{condition-5}, and 
necessarily $\kappa<\epsilon\calL_{\rm +e}^2/\calL_{\rm +p}^2$ and so $\Delta P<0$ holds. 
The charge acquired by the black hole $\qE$ can become the smallest value $\qmin$ if $T_{\rm e}$ 
is less than $\kappa_{\rm min}T_{\rm p}$. By contrast, if $\kappa>\kappa_{\rm max}$, 
or equivalently, $T_{\rm e}>\kappa_{\max}\Tp$ holds, the charge of the black hole $\qE$ will approach $\qmax$.

\begin{figure}[!h]
\centering\includegraphics[width=10cm]{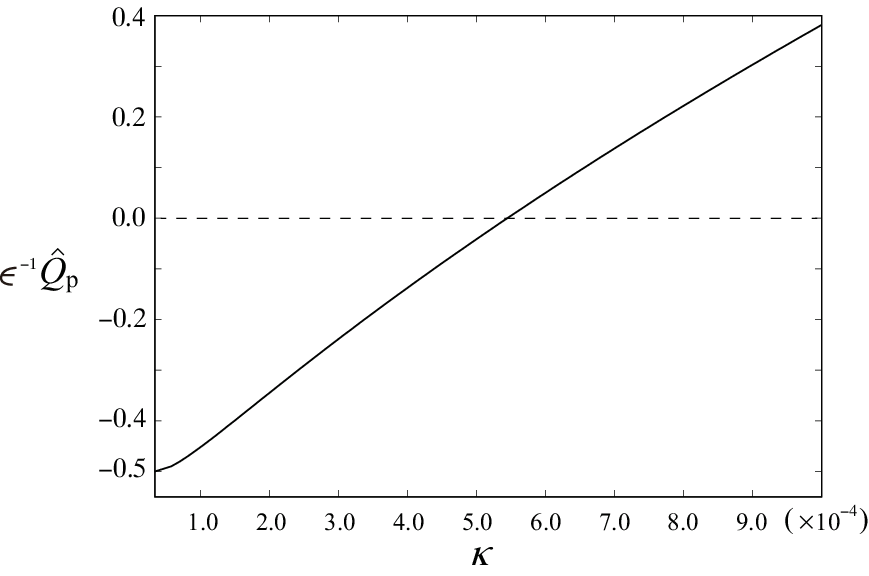}
\caption{ The same as Fig.~\ref{fig:kappa-charge} but  $\epsilon^{-1}\hcalQp=-\hcalQe$ is depicted in the domain $\kappa_{\rm min}<\kappa<10^{-3}$. }
\label{fig:kappa-charge-small}
\end{figure}

If the temperature of the proton and that of the electron are identical to each other, $\kappa=1\gg\epsilon$ holds, and hence  
the black hole might acquire the positive charge. 
By numerically solving Eq.~\eqref{condition-c} with $\kappa=1$, we have $\hcalQp=3.79\times10^{-2}$ and can estimate the charge 
$\hqE$ acquired by a non-rotating black hole as
\begin{equation}
\hqE=3.79\times10^{-2}\times \frac{8\pi\epzero GM\mpr}{e}
\simeq4.67\times 10^7\left(\frac{M}{4.0\times10^6M_\odot}\right){\rm C}.
\label{equi-temp-charge}
\end{equation}
In the astrophysical situations we are interested in, the inequality $\Te>\Tp$ often holds and hence many black holes in our Universe might be positively electrified.

\subsection{The case of $\calQ>1/2$}

In this case, $G_+$ has a maximum even in the case of $\calL^2=0$. Since the maximum value of $G_+$ 
is an increasing function of $\calL^2$ from Eq.~\eqref{delG-delL}, 
the maximum value of $G_+$ with $\calL^2=0$ is the smallest. In the case of $\calL^2=0$, 
the radius $\Rmax$ is given as
\begin{equation}
\Rmax(0;\calQ)=\frac{4\calQ^2}{4\calQ^2-1}.
\end{equation}
The maximum value of $G_+(R;0,\calQ)$ is given as
\begin{equation}
G_+\bigl(\Rmax(0;\calQ);0,\calQ\bigr)=\calQ+\frac{1}{4\calQ}.
\end{equation}
We can easily see that the RHS of this equation is an increasing function of $\calQ$. Hence, we have
\begin{equation}
\left.G_+\bigl(\Rmax(0;\calQ);0,\calQ\bigr)\right|_{\calQ>1/2}> \left.G_+\bigl(\Rmax(0;\calQ);0,\calQ\bigr)\right|_{\calQ=1/2}=1.
\end{equation}
Thus, we have $G_+\bigl(\Rmax(\calL,\calQ);\calL,\calQ\bigr)>1$, or equivalently,
\begin{equation}
\Lambda(\calL^2;\calQ)>0.
\end{equation}
We can easily see that the following equation holds. 
\begin{equation}
\Lambda(0;\calQ)=\left(\calQ-\frac{1}{4\calQ}\right)^2>0.
\end{equation}
By using Eq.~\eqref{delG-delL}, we obtain 
\begin{align}
\left.\frac{\partial\Lambda\left(\calL^2;\calQ\right)}{\partial\calL^2}\right|_{\calL^2=0}=\frac{\left(4\calQ^2+1\right)\left(4\calQ^2-1\right)^2}{128\calQ^6}. 
\end{align}
Hence, we have
\begin{equation}
\Lambda\left(\calL^2;\calQ\right)=\left(\frac{4\calQ^2-1}{4\calQ}\right)^2\left[1+\frac{4\calQ^2+1}{8\calQ^4}\calL^2\right]+{\cal O}\left(\calL^4\right).
\end{equation}

The probability $P$ for the charged test particle with $\calQ>1/2$ to enter the black hole is given by
\begin{align}
P&=\int_0^{2\pi}d\Theta\int_0^\infty d\vomg \vomg \int_{-\infty}^{-c\sqrt{\Lambda(R^2\vomg^2/c^2;\calQ)}}dv_{(r)}
\left(\frac{m}{2\pi \kB T}\right)^{3\over2}\exp\left[-\frac{m} {2\kB T}\left(v_{(r)}^2+\vomg^2\right)\right] \nonumber \\
&=\frac{mc^2}{4\kB TR^2}\int_0^\infty d\calL^2 \exp\left(-\frac{mc^2} {2\kB T R^2}\calL^2\right)
{\rm erfc}\left[\sqrt{\frac{mc^2}{2\kB T}\Lambda\left(\calL^2;\calQ\right)}\right] \nonumber \\
&< \frac{mc^2}{4\kB TR^2}\int_0^\infty d\calL^2 
\sqrt{\frac{2\kB T}{\pi mc^2\Lambda\left(\calL^2;\calQ\right)}}\exp\left[-\frac{mc^2}{2\kB T}\Lambda\left(\calL^2;\calQ\right)\right],\label{P-c}
\end{align}
where erfc$[x]$ is the Gauss's complementary error function, and we have used the following inequalities
\begin{align}
\exp\left(-\frac{mc^2} {2\kB T R^2}\calL^2\right)&<1, \\
\int_z^\infty e^{-x^2}dx-\frac{e^{-z^2}}{2z}=-\int_z^\infty \frac{e^{-x^2}}{2x^2}dx&<0
\end{align}
in the last inequality. Since $\Lambda$ is a monotonically increasing function of $\calL^2$, we have 
\begin{align}
P&<\frac{1}{R^2}\sqrt{\frac{mc^2}{8\pi\kB T \Lambda(0;\calQ)}}\int_0^\infty d\calL^2 \exp\left[-\frac{mc^2}{2\kB T}\Lambda\left(\calL^2;\calQ\right)\right] \nonumber \\
&=\frac{1}{R^2}\sqrt{\frac{mc^2}{8\pi\kB T \Lambda(0;\calQ)}}
\int_{\Lambda(0;\calQ)}^\infty d\Lambda \left(\frac{\partial\Lambda}{\partial \calL^2}\right)^{-1}
\exp\left(-\frac{mc^2} {2\kB T}\Lambda\right)\nonumber \\
&\simeq\frac{1}{R^2}\sqrt{\frac{mc^2}{8\pi\kB T \Lambda(0;\calQ)}}
\left(\left.\frac{\partial\Lambda}{\partial \calL^2}\right|_{\calL^2=0}\right)^{-1}
\exp\left(-\frac{mc^2} {2\kB T}\Lambda\left(0;\calQ\right)\right)\nonumber \\
&=\frac{256 \calQ^7}{\left(4\calQ^2+1\right)\left(4\calQ^2-1\right)^3R^2}\sqrt{\frac{2\kB T}{\pi mc^2}}
\exp\left[-\frac{mc^2}{2\kB T}\left(\frac{4\calQ^2-1}{4\calQ}\right)^2\right],
\end{align}
where in the third line with an approximate equality, we have evaluated the integral by using 
\begin{equation}
\lim_{z\rightarrow\infty}e^z\int_z^\infty  F(x) e^{-x}dx=\lim_{z\rightarrow\infty}F(z)
\end{equation}
because of the assumption $mc^2/\kB T\gg1$. 
We should note that this estimate is valid only if the following inequality holds,
\begin{equation}
\calQ-\frac{1}{2}\gg \left(\frac{\kB T}{mc^2}\right)^{1\over2}. \label{validity-condition}
\end{equation} 
Very careful limiting procedure is necessary to evaluate $P$ for $\calQ\rightarrow1/2+0$. 

If Eq.~\eqref{validity-condition} holds, the upper bound of the probability for the test charged particle to enter the black hole is suppressed 
by the small factor 
\begin{equation}
\frac{512\calQ^7}{\sqrt{\pi}\calL_+^2\left(4\calQ^2+1\right)\left(4\calQ^2-1\right)^3}
\left(\frac{2\kB T}{mc^2}\right)^{3\over2}\exp\left[-\frac{mc^2}{2\kB T}\left(\frac{4\calQ^2-1}{4\calQ}\right)^2\right],
\end{equation}
compared to Eq.~\eqref{P-charged} for $\calQ \leq1/2$. Hence, we may say that the charge acquired by a non-rotating black hole 
is bounded by $\qmin\leq q \leq \qmax$ in the case of the non-relativistic velocity distribution $\kB T/mc^2\ll1$, 
where $\qmin$ and $\qmax$ have been defined as Eqs.~\eqref{minimal-c} and \eqref{maximal-c}.

\section{Proton and electron on ISCO}\label{sec:ISCO}

The ISCO is important to estimate the angular diameter of a black hole shadow and luminosity accompanied 
by the accretion process of matter to a black hole. We study the ISCO of a proton and an electron around 
a non-rotating charged black hole. 

Note that $\calE$ cannot be larger than or equal to one for the stable circular orbit. 
We rewrite the effective potential defined as Eq.~\eqref{potential-c} in the form 
\begin{equation}
U(R;\calE,\calL^2,\calQ)=1-\calE^2-\frac{1-2\calQ\calE}{R}+\frac{\calL^2-\calQ^2}{R^2}-\frac{\calL^2}{R^3}.
\end{equation}
The radial coordinate of the stable circular orbit is a local minimum and zero point of the effective potential;
\begin{equation}
U=0,~~~~\frac{\partial U}{\partial R}=0~~~~{\rm and}~~~~\frac{\partial^2 U}{\partial R^2}>0. \label{SCO}
\end{equation}
The second equation of Eq.~\eqref{SCO} is equivalent to a quadratic equation with respect to $R$ as 
\begin{equation}
\left(1-2\calQ\calE\right)R^2-2\left(\calL^2-\calQ^2\right)R+3\calL^2=0. \label{extrema}
\end{equation}
Since $\partial U/\partial R\rightarrow+\infty$ for $R\rightarrow0$,  the stable circular orbit 
exists only if there are two extrema in the domain of $R>0$; if there is only one extremum of $U$ in the domain $R>0$, 
except for  the double root case, 
it is necessarily a maximum which should correspond to the unstable circular orbit. 
Hence the stable circular orbit exists only if Eq.~\eqref{extrema} has two positive roots,  
or in other words, only if
\begin{equation}
1-2\calQ\calE>0~~~~{\rm and}~~~~\calL^2-\calQ^2>0 \label{existence}
\end{equation}
hold. 

The ISCO is a circular orbit whose radius is a double root of Eq.~\eqref{extrema}, i.e., a marginally stable circular orbit. 
The dimensionless specific energy $\calE$ and the dimensionless specific angular momentum $\calL$ of the ISCO 
should lead to the vanishing discriminant of Eq.~\eqref{extrema};
\begin{equation}
\left(\calL^2-\calQ^2\right)^2-3\calL^2(1-2\calQ \calE)=0, 
\end{equation}
or equivalently,
\begin{equation}
1-2\calQ \calE=\frac{\left(\calL^2-\calQ^2\right)^2}{3\calL^2}. \label{discriminant-c}
\end{equation}
Then, solving Eq.~\eqref{extrema}, the ISCO radius is given by
\begin{align}
R:=\frac{\calL^2-\calQ^2}{1-2\calQ\calE} 
=\frac{3\calL^2}{\calL^2-\calQ^2} 
=3\left(1+\frac{\calQ^2}{\calL^2-\calQ^2}\right) 
>3, \label{R-isco-0}
\end{align}
where we have used Eq.~\eqref{discriminant-c} in the second equality. Note that the positivity of the ISCO radius leads to $\calL^2-\calQ^2>0$ and 
so the final inequality is derived. This result implies that the ISCO radius of a charged test particle is necessarily larger than that of a neutral test 
particle, $R=3$, or equivalently, $r=3\rg$, despite of the sign of the electric charge of the particle as long as  the black hole has the electric charge. 
This result contradicts the result obtained by Zaja\v{c}ek {\it et al.} \cite{Zajacek-etal}. Since Zaja\v{c}ek {\it et al.} 
did not explicitly show the definition of the ISCO, 
their definition might be different from ours. 

By substituting Eqs.~\eqref{discriminant-c} and \eqref{R-isco-0} into the first equation of Eq.~\eqref{SCO}, we obtain
\begin{equation}
\calE^2=1-\frac{\left(\calL^2-\calQ^2\right)^3}{27\calL^4}. \label{E-isco-0}
\end{equation}
Thus, the dimensionless orbital radius, $\Risco$, and the dimensionless specific energy, $\Eisco$, of the charged test 
particle on the ISCO are given as functions of $\calL^2$ and $\calQ^2$ in the form,
\begin{align}
\Risco&=\frac{3\calL^2}{\calL^2-\calQ^2}, \label{R-isco} \\
\Eisco&=\sqrt{1-\frac{\left(\calL^2-\calQ^2\right)^3}{27\calL^4}}. \label{E-isco}
\end{align}

By solving Eq.~\eqref{discriminant-c} with respect to $\calE$, we have
\begin{equation}
\calE=\frac{1}{6\calQ\calL^2}\left[3\calL^2-\left(\calL^2-\calQ^2\right)^2\right]. \label{E-isco-1}
\end{equation}
The dimensionless specific angular momentum of the ISCO particle is determined by the equation derived by substituting 
Eq.~\eqref{E-isco} into the L.H.S. of Eq.~\eqref{E-isco-1}; 
\begin{equation}
\sqrt{3}\left[3\calL^2-\left(\calL^2-\calQ^2\right)^2\right]
-2\calQ\sqrt{27\calL^4-\left(\calL^2-\calQ^2\right)^3}=0. \label{calL-eq}
\end{equation}
We numerically solve this equation to obtain $\calL$ for $-\epsilon/2<\calQ<1/2$ by the Newton method 
and substitute the results into Eqs.~\eqref{R-isco} and \eqref{E-isco} in order to obtain $\Risco$ and $\Eisco$.

\begin{figure}[!h]
\centering\includegraphics[width=9cm]{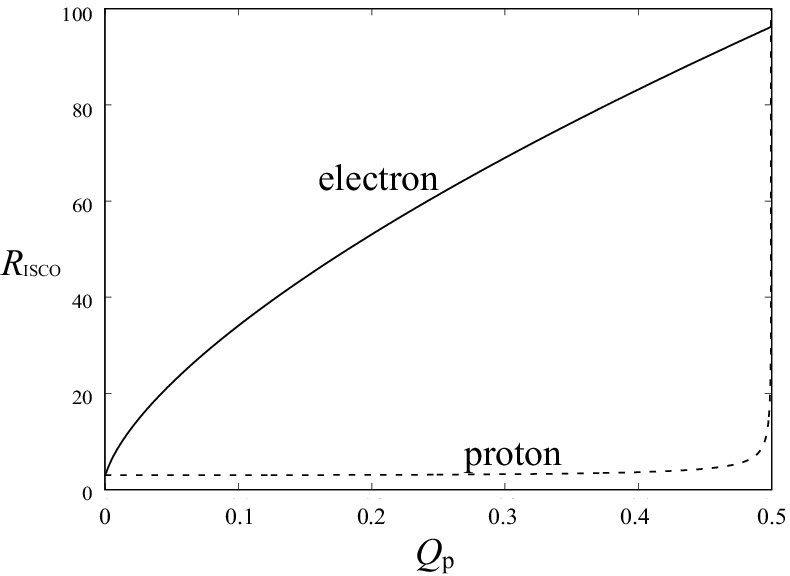}
\caption{ The orbital radii of the proton and the electron on ISCO are depicted as functions of $\calQp$ which is equal to $-\epsilon\calQe$. 
 }
\label{fig:ISCO-radius}
\end{figure}

\begin{figure}[!h]
\centering\includegraphics[width=9cm]{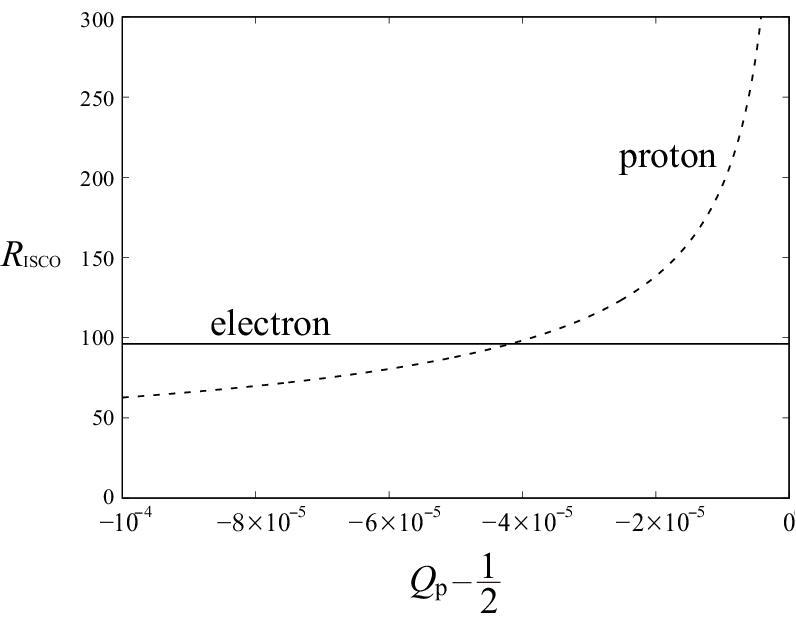}
\caption{ The same as Fig.~\ref{fig:ISCO-radius} but  in the domain $1/2-10^{-4}\leq\calQp\leq1/2$. 
}
\label{fig:ISCO-radius-large}
\end{figure}

\begin{figure}[!h]
\centering\includegraphics[width=9cm]{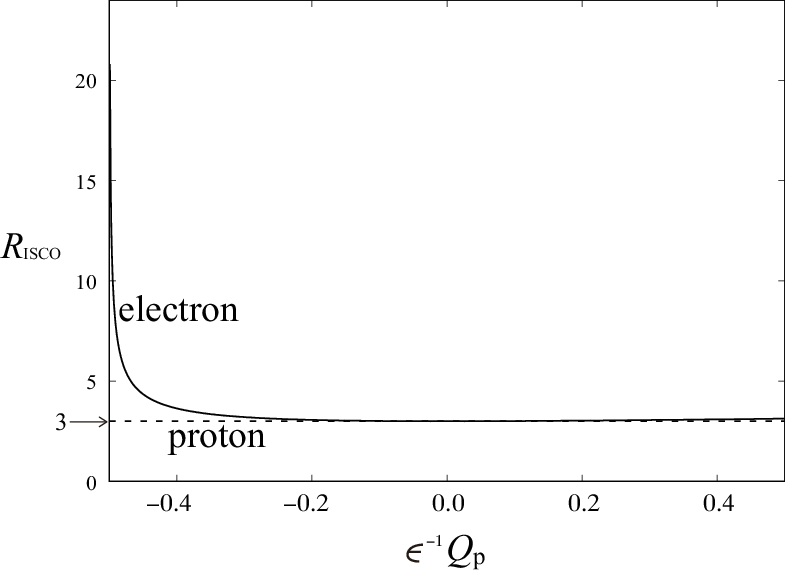}
\caption{ The same as Fig.~\ref{fig:ISCO-radius} but  as a function of $\epsilon^{-1}\calQp$ in the domain $[-1/2,1/2]$. 
Note that $\epsilon^{-1}\calQp=-\calQe$ holds. }
\label{fig:ISCO-radius-small}
\end{figure}

The dimensionless ISCO radii, $\Risco$, of a proton and an electron are depicted as functions of $\calQp$ 
in the domain $-\epsilon/2\leq\calQp\leq1/2$ 
in Fig.~\ref{fig:ISCO-radius}, 
in the domain $-10^{-4}\leq\calQp-1/2\leq0$ 
in Fig.~\ref{fig:ISCO-radius-large}, 
and as a function of $\epsilon^{-1}\calQp$ in the domain $-1/2\leq\epsilon^{-1}\calQp\leq1/2$ in Fig.~\ref{fig:ISCO-radius-small}. 
Note that  $\epsilon^{-1}\calQp$ is equal to $-\calQe$. 

From these figures, we find that the ISCO radius of the proton is nearly equal to $3\rg$ for $\calQp\lesssim0.48$ but 
grows rapidly in $\calQp\gtrsim 0.48$ and diverges at $\calQp=1/2$. By contrast, the ISCO radius of the electron is  
larger than that of the proton except in the domain very close to $\calQp=1/2$, while it diverges at $\calQe=-\epsilon^{-1}\calQp=1/2$. 
If $\calQp$ is very close to $1/2$, ISCO radii of both the proton and electron 
are much larger than that of the neutral particle. Thus, the structure of an accretion disk composed of the electron-proton plasma 
around an electrically charged black hole with $\qE\simeq\qmax$ might be very different from that around a neutral black hole.

\begin{figure}[!h]
\centering\includegraphics[width=10cm]{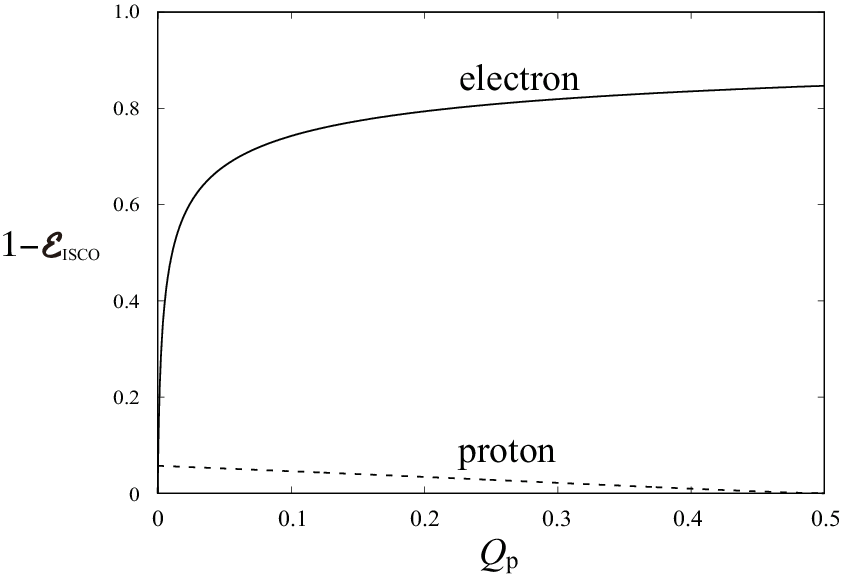}
\caption{ The specific energies of the proton and the electron on ISCO are depicted as functions of $\calQp$ which is equal to $-\epsilon\calQe$. 
 }
\label{fig:ISCO-energy}
\end{figure}

\begin{figure}[!h]
\centering\includegraphics[width=10cm]{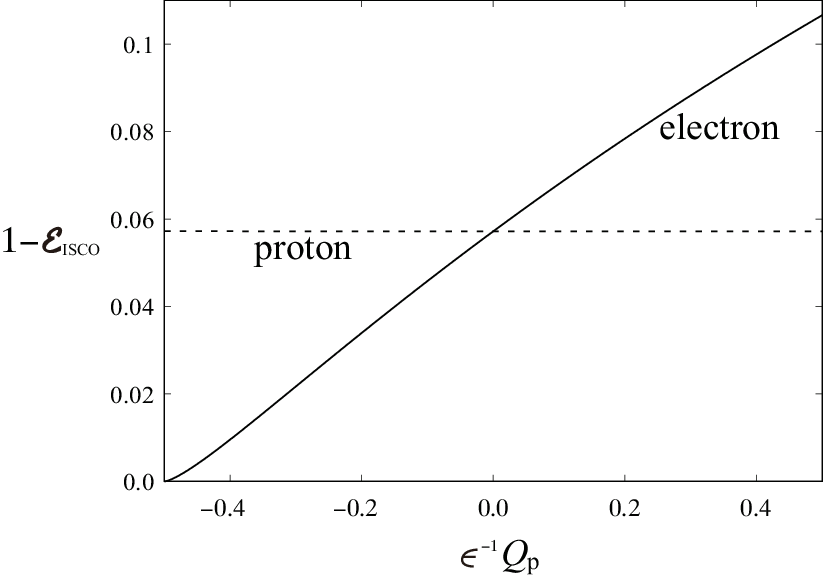}
\caption{ The same as Fig.~\ref{fig:ISCO-energy} but  as a function of $\epsilon^{-1}\calQp$ in the domain $[-1/2,1/2]$. 
Note that $\epsilon^{-1}\calQp$ is equal to $-\calQe$. }
\label{fig:ISCO-energy-small}
\end{figure}

Since the dimensionless energy $\calE$ of the particle on the circular orbit with infinite radius is one, $1-\Eisco$ is equal to the ratio of the released energy 
by the accretion of this particle to its rest mass energy, and hence we depict $1-\Eisco$ instead of $\Eisco$ itself as functions of $\calQp$  in the domain 
$-\epsilon/2\leq\calQp\leq1/2$ in Fig.~\ref{fig:ISCO-energy} and as a function of $\epsilon^{-1}\calQp$ in the domain $-1/2\leq\epsilon^{-1}\calQp\leq1/2$ 
in Fig.~\ref{fig:ISCO-energy-small}. The quantity $1-\Eisco$ of the proton for small $\calQp$ is almost equal to the value of the neutral test particle but 
in the limit to the maximal value $\calQp\rightarrow1/2$, it vanishes. Hence, the released energy due to the accretion of protons around a black hole 
with $\qE\simeq\qmax$ might be much less than that of neutral test particles. By contrast, $1-\Eisco$ of the electron can be much larger than that of the 
neutral test particle. However it vanishes in the limit of $\calQe=-\epsilon^{-1}\calQp\rightarrow1/2$.

The following two cases might be of perticular interest. 

\subsection{Equi-temperature $\kappa=1$}

For a proton, the dimensionless ISCO radius $\Risco$ and the specific energy, $1-\Eisco$, released by its accretion
are numerically obtained as follows.
\begin{align}
\Risco-3&=1.55\times10^{-3}, \\
1-\Eisco&=5.29\times10^{-2}.
\end{align}
The radius $\Risco$ and $1-\Eisco$ of the proton on the ISCO
are almost equal to the values of a neutral test particle, 3 and $1-\sqrt{8/9}\simeq5.72\times10^{-2}$, respectively. 
These results come from the smallness of $\calQp$ which is equal to $3.79\times10^{-2}$. 

By contrast, for an electron, we have  
\begin{align}
\Risco&=18.7, \\
1-\Eisco&=0.652.
\end{align}
These values very different form those of a neutral test particle come from the large value of $|\calQe|$ which is equal to $\epsilon^{-1}\calQp=69.6$. 

These results imply that if the plasma density is so small that the Coulomb force between electrons and protons are negligible, 
the number of protons near the ISCO of the neutral test particles will be much larger than that of electrons in the equi-temperature case. 

\subsection{Equi-ISCO-radius}

Another case which might be important is that the ISCO radius of the proton is equal to that of the electron. 
From Figs.~\ref{fig:ISCO-radius} and \ref{fig:ISCO-radius-large}, 
we can see that it is realized at $\calQp$ very close to 1/2 except in the trivial case $\calQp=0$. 
The values of $\calQp$ and $\Risco$ in the case of equi-ISCO-radius 
are numerically obtained as
\begin{align}
\frac{1}{2}-\calQp&=4.17\times 10^{-5}, \\
\Risco&=96.2.
\end{align}
In this case, the released specific energy by accretion of the proton is given as
\begin{equation}
1-\Eisco=1.45\times10^{-7} ,
\end{equation}
 whereas we have, for the electron, 
 \begin{equation}
 1-\Eisco=0.847.
 \end{equation}
 Since $\calQp$ is less than and very close to $1/2$, $\kappa$ 
 is less than and very close to $\kappa_{\rm max}=2.01\times10^3$. The electron temperature should be much larger than 
 that of the proton so that the ISCO radii of the proton and electron are identical to each other 
 if the consideration in the previous section is applied to this case.  
 
We can see from these results that $85\%$ of the rest mass energy of the electron may be released though its accretion, 
whereas only about $10^{-5}\%$ of that of the proton may be released.  
The ratio of the energy released by the accretion of a pair of a proton and electron to that of a neutral hydrogen is 
estimated as 
\begin{equation}
\frac{1.5\times10^{-7}\mpr c^2+0.85 \mel c^2}{\left(1-\sqrt{8/9}\right)\left(\mpr+\mel\right)c^2}=8.1\times10^{-3}.
\end{equation} 
The energy extraction through the accretion process of plasma might be much more inefficient than that of the neutral matter if the 
black hole has the electric charge so as to be equi-ISCO radius.

\section{Summary}\label{sec:Discussion}

We studied the electrification of a non-rotating black hole by the accretion of protons and electrons with thermal 
velocity distributions through the analysis of their motion by assuming them to be charged test particles 
in the Schwarzschild spacetime and investigated the effects of the charge acquired by the black hole on the motion of 
a proton and an electron through the investigation of their ISCO. 

We showed that, assuming that the initial velocities of the protons and the electrons follow the Maxwell distribution,
the  difference appears between the probability of a proton to enter a non-rotating black hole and that of an electron  
due to the difference of mass between a proton and an electron and due to the temperature difference.  
This result will lead to the electrification of a non-rotating black hole if the black hole is surrounded by plasma so diffuse that the Coulomb  
force between protons and electrons is smaller than the tidal force between them due to the black hole. 
We also estimated the maximal electric charge acquired by the black hole as Eq.~\eqref{maximal-c}. 
We saw, by an order estimate, that the Coulomb force between protons and electrons should be taken into account in the case of 
supermassive black holes like as Srg A*, whereas it might be negligible in the case of a stellar mass black hole.  

Although, exactly speaking, a non-rotating charged black hole is described by the Reissner-Nordstr\"{o}m solution, the charge parameter 
of the Reissner-Nordstr\"{o}m black hole electrified in the situation considered in this paper is so small that the spacetime geometry 
is well described by the Schwarzschild solution. Thus, the motion of a neutral test particle is not affected by the electrification of the 
non-rotating black hole. By contrast, the motion of a charged test particle can be largely affected. 
We studied the ISCO of a proton and an electron in the spacetime with a non-rotating charged black hole 
and showed that there can be large differences between the proton, the electron and a neutral particle if the black hole acquires electric charge 
very close to the maximal value given by \eqref{maximal-c}. This result implies that the efficiency of the energy extraction 
due to the accretion of plasma to the charged black hole might be much less than that of neutral matter. 
However, the Coulomb force between protons and electrons should be taken into account in the accretion process. 
This subject is out of scope of this paper and will be discussed elsewhere in future.

\section*{Acknowledgments}
We are grateful to colleagues, especially K. Ueda and K. Sueto, in the astrophysics and gravity group in Osaka Metropolitan University. 
This work was supported by JSPS KAKENHI Grants No. JP21K03557 (K.N.), No. JP21H05189 (H.Y.), No. JP22H01220 (H.Y.), 
and MEXT Promotion of Distinctive Joint Research Center Program JPMXP0723833165 (K.N. and H.Y.).

\appendix
\section{Orthonormal frame associated with a static observer}\label{orthonormal}

In this paper, we adopt an orthonormal frame associated with a static observer, which is defined as
\begin{align}
e_{(0)\mu}&=\left(-\sqrt{f(r)},0,0,0\right), \label{et-def} \\
e_{(1)\mu}&=\left(0,\frac{1}{\sqrt{f(r)}},0,0\right), \\
e_{(2)\mu}&=\left(0,0,r,0\right),\\
e_{(3)\mu}&=\left(0,0,0,r\sin\theta\right) \label{ep-def}
\end{align}
and 
\begin{equation}
e^{(\mu)}{}_\alpha=\sum_{\nu=0}^3 \eta^{(\mu)(\nu)}e_{(\nu)\alpha},
\end{equation}
where $\eta^{(\mu)(\nu)}={\rm diag}[-1,1,1,1]$ is the Minkowskian inverse metric. 
Then the components of the 3-velocity of a test particle with the 4-velocity $u^\mu$ 
with respect to the orthonormal frame \eqref{et-def}--\eqref{ep-def} are defined as
\begin{equation}
\left(v_{(r)},v_{(\theta)},v_{(\varphi)}\right):=\left(u_{(1)},u_{(2)},u_{(3)}\right),
\end{equation}
where $u^{(\mu)}=e^{(\mu)}{}_\alpha u^\alpha$ and $u_{(\mu)}=e_{(\mu)\alpha}u^\alpha$. 
Then we obtain Eqs.~\eqref{vr-def}--\eqref{vp-def}. 

The momentum distribution function $f_{\rm m}$ of the relativistic ideal gas in the thermal equilibrium with the temperature $T$ is given as
\begin{equation}
f_{\rm m}(\bm{p})=\frac{N}{\exp\left(\dfrac{c\sqrt{\bm{p}^2+m^2c^2}}{\kB T}\right)\pm1},
\end{equation}
where $N$ is the normalization constant, 
$\bm{p}$ is the spatial components of the 4-momentum, and the plus sign is taken for the fermion, whereas the minus sign is taken for 
the boson. In the non-relativistic situation, $\bm{p}^2\ll m^2c^2$, we have Eq.~\eqref{Maxwell}.

\section{Behavior of $G_+(R;\calL,\calQ)$}\label{G-plus}

Extrema of the function $G_+$ are obtained as roots of the equation 
\begin{equation}
\frac{\partial G_+(R;\calL,\calQ)}{\partial R}=0, \label{G-extrema} 
\end{equation}
where introducing the following functions
\begin{align}
D_1(R;\calL)&=R^2-2\calL^2 R+3\calL^2, \\
D_2(R;\calL,\calQ)&=-2\calQ\sqrt{R\left(R-1\right)\left(R^2+\calL^2\right)},
\end{align}
the derivative of $G_+$ with respect to $R$ is written as
\begin{equation}
\frac{\partial G_+(R;\calL,\calQ)}{\partial R}=\frac{D_1(R;\calL)+D_2(R;\calL,\calQ)}{2R^2\sqrt{R(R-1)\left(R^2+\calL^2\right)}}.
\end{equation}

Then the dimensionless radius $R=\Rmax$ at the extremum of $G_+$ is a positive root of $D_1+D_2=0$, or equivalently, 
\begin{equation}
R^2-2\calL^2 R+3\calL^2-2\calQ\sqrt{R\left(R-1\right)\left(R^2+\calL^2\right)}=0.
\end{equation}
We can numerically confirm that $\Rmax$ has a finite limiting value for $\calL^2\rightarrow\infty$. Hence in the limit of $\calL^2\gg1$, 
this equation takes the from
\begin{equation}
-2\calL^2 R+3\calL^2-{\cal O}\left(|\calL|\right)=0.
\end{equation}
Hence, for $\calL^2\gg1$, we have
\begin{equation}
\Rmax=\frac{3}{2}+{\cal O}\left(|\calL|^{-1}\right).
\end{equation}
By using this result, we have 
\begin{equation}
\Lambda\left(\calL^2;\calQ\right)=\frac{4}{27}\calL^2\left[1+{\cal O}\left(|\calL|^{-1}\right)\right].  
\end{equation}

\section{End of the electrification}\label{End}

We consider the accretion of plasma composed of protons and electrons to a non-rotating 
black hole in the situation studied in Sec.~\ref{Q-le-half}.  
In the case that $N$ plasma particles fall into the black hole, the number of falling protons is equal to $N\Pp/(\Pp+\Pe)$, whereas that 
of falling electrons is equal to $N\Pe/(\Pp+\Pe)$. The increment of the electric charge of the black hole, $\Delta\qE$, is given as
\begin{equation}
\Delta\qE=\frac{Ne}{\Pp+\Pe}\left(\Pp-\Pe\right),
\end{equation}
whereas the increment of the mass of the black hole, $\Delta M_{\rm BH}$, is given as
\begin{equation}
\Delta M_{\rm BH}=\frac{N\mpr}{\Pp+\Pe}\left(\Pp+\epsilon\Pe\right).
\end{equation}
Note that $\Delta M_{\rm BH}$ should be non-negative. 
Hence, the increment of the charge parameter $\calQp$ of a proton, which is denoted by $\Delta\calQp$, is given as
\begin{align}
\Delta\calQp&=\calQp\left[\left(1+\frac{\Delta\qE}{\qE}\right)\left(1+\frac{\Delta M_{\rm BH}}{M_{\rm BH}}\right)^{-1}-1\right] \nonumber\\
&= \calQp\left(\frac{\Delta\qE}{\qE}-\frac{\Delta M_{\rm BH}}{M_{\rm BH}}\right)\left(1+\frac{\Delta M_{\rm BH}}{M_{\rm BH}}\right)^{-1} \nonumber \\
&=\frac{Ne\calQp}{\qE}\frac{\Pp+\epsilon\Pe}{\Pp+\Pe}
\left[\frac{\Pp-\Pe}{\Pp+\epsilon\Pe}-\frac{\qE}{M_{\rm BH}}\left(\frac{e}{\mpr}\right)^{-1}\right]
\left(1+\frac{\Delta M_{\rm BH}}{M_{\rm BH}}\right)^{-1}, \label{DelQp}
\end{align}
where $M_{\rm BH}$ is the mass of the black hole. 

The end of the electrification of the black hole corresponds to 
$\Delta \calQp=0$ which leads to
\begin{equation}
\frac{\Pp-\Pe}{\Pp+\epsilon\Pe}=\frac{\qE}{M_{\rm BH}}\left(\frac{e}{\mpr}\right)^{-1}
=\frac{2Q_{\rm BH}}{\rg}\left(\frac{m_{\rm pl}}{\mpr}\sqrt{\alpha}\right)^{-1},
\end{equation}
where $Q_{\rm BH}$ is the charge parameter of the Reissner-Nordstr\"{o}m black hole with the electric charge $\qE$, and  
$m_{\rm pl}$ is the Planck mass ($\simeq 1.3\times10^{19}\mpr$).  
Since $\left|Q_{\rm BH}\right|/\rg < 1$ must hold in the case of our interest, the following inequality 
\begin{equation}
\frac{\left|\Pp-\Pe\right|}{\Pp}\ll1 \label{end-condition}
\end{equation}
should hold, if $\Delta\calQp=0$ holds. Conversely, if $\Pp-\Pe=0$ holds, we can see from Eq.~\eqref{DelQp} that 
$\left|\Delta\calQp/\calQp\right|\ll1$ holds, as long as $Ne/\qE\ll m_{\rm pl}/\mpr$ is satisfied. Hence, Eq.~\eqref{condition-c} is a very good approximation 
of the necessary and sufficient condition for the end of the electrification of the black hole.

\end{document}